\documentclass{PoS}

\title{$\alpha_s$ in 2016 from the (revised) ALEPH data for $\tau$ decay}

\ShortTitle{$\alpha_s$ in 2016 from $\tau$ decay}

\author{Diogo Boito\\
       Instituto de F\'{\i}sica de S{\~a}o Carlos, Universidade de S{\~a}o Paulo\\
CP 369, 13570-970, S{\~a}o Carlos, SP, Brazil\\
       }

       \author{Maarten Golterman\\
      Dept. of Physics and Astronomy, San Francisco State University\\ San Francisco, CA 94132, USA\\
       }

  \author{Kim Maltman\\
       Dept. of Mathematics and Statistics, York Univ.\\Toronto, ON Canada M3J~1P3\\
        CSSM, University of Adelaide\\ Adelaide, SA~5005, Australia\\
       }

\author{\speaker{Santiago Peris}
\\
        Dept. of Physics and IFAE-BIST, Univ.  Aut\`{o}noma de Barcelona\\
        E-08193 Bellaterra, Barcelona, Spain\\
        E-mail: \email{peris@ifae.es}}

\abstract{We summarize a comparison of the two strategies which are currently available in the literature for determining the value of $\alpha_s(m_\tau)$. We will refer to these as the truncated Operator Product Expansion model and the Duality Violation model. After describing the main features of both approaches, we explain why the former fails to pass crucial tests. The latter, on the other hand, passes all the tests known up to date and, therefore, should be currently considered the only reliable method.}

\FullConference{38th International Conference on High Energy Physics\\
		3-10 August 2016\\
		Chicago, USA}

\begin{document}

The $\tau$ is the only lepton capable of decaying into hadrons. This makes it
a potentially clean experimental environment to study QCD and, particularly,
to determine the strong coupling $\alpha_s$. Furthermore, its mass
$m_{\tau}\simeq 1.8\, \mathrm{GeV}$ is both large enough to allow the use of
perturbation theory (supplemented by nonperturbative corrections), and small
enough to potentially compete in precision with other
determinations of $\alpha_s$, carried out at much higher energies such
as the $Z$ mass. This is because the relative error in $\alpha_s$ at the $Z$
mass, $\epsilon(M_Z)$, gets squeezed from its value at the $\tau$ scale by
$\epsilon(M_Z)\simeq \epsilon(m_{\tau})
(\alpha_s(M_Z)/\alpha_s(m_\tau))$. Since no lunch is free, however, the price
to pay at the $\tau$ scale is that
nonperturbative corrections are non-negligible,
and it is absolutely essential to bring them under good theoretical control.
Otherwise, the determination of $\alpha_s$ becomes unreliable.

In 1992, Braaten, Narison and Pich \cite{BNP}, followed by LeDiberder and Pich
\cite{LDP}, building on previous work \cite{Shankar}-\cite{Braaten88}, showed
how Perturbation Theory (PT) and the Operator Product Expansion (OPE)
could potentially be used to produce a precise determination of
$\alpha_s(m_\tau)$. This method has been the traditional choice
until recently, and has been applied in a series of analyses based on
the hadronic data collected in the OPAL \cite{OPAL} and the
ALEPH \cite{ALEPH,ALEPH13} experiments.\footnote{The latest ALEPH data were corrected in Ref.~\cite{ALEPH13} after an error in its covariance matrices was detected in Ref.~\cite{Diogo}.}

The error originally estimated for $\alpha_s(m_\tau)$ in 1992
ranged from $10\%-30\%$ \cite{BNP}; that currently claimed in
Ref.~\cite{ALEPH13} is roughly $2\%$ (in CIPT).
Of course, most of the improvement in precision is due to the more accurate
data presently available. This improvement in experimental precision,
however, should also be accompanied by an improvement in the theoretical
description used to analyze these data.  The theoretical framework used
in these traditional analyses has, however, remained largely the
unmodified version of the original one from Ref. \cite{LDP}.

This motivated us in a series of papers \cite{Cata0}-\cite{Boito3} to take
a fresh look at the assumptions underlying these traditional
determinations of $\alpha_s(m_\tau)$. The outcome of these investigations is a
new method to analyze the hadronic $\tau$ data. This led to a new,
significantly lower result for $\alpha_s(m_\tau)$ in Ref.~\cite{us}.
This calls, of course, for a closer investigation.

Here we would like to point out some of the main differences between the two
methods. In disagreement with Ref.~\cite{Pich} our conclusion will be that
the method discussed in Ref.~\cite{Pich} has fundamental flaws and,
therefore, that the method of Ref.~\cite{us} is the only reliable one
currently available.
Due to space limitations, only a brief summary of our results will be presented. The reader is referred to Ref. \cite{newpaper} for a more detailed account.

All analyses start with the following equation \cite{BNP}-\cite{Braaten88}:
\begin{eqnarray}
\label{sumrule}
&&\hspace{-1cm}\frac{1}{s_0}\int_0^{s_0}ds\,w(s/s_0)\,\rho^{(1+0)}_{V/A}(s) \\
&&=
-\frac{1}{2\pi is_0}\oint_{|s|=s_0}
ds\,w(s/s_0)\,\Pi^{(1+0)}_{{\rm OPE},V/A}(s)-\frac{1}{s_0}\,
\int_{s_0}^\infty ds\,w(s/s_0)\,\frac{1}{\pi}\,\mbox{Im}\,
\Delta_{V/A}(s)\ ,\nonumber
\end{eqnarray}
where $w(x)$ is a convenient polynomial, $\rho^{(1+0)}_{V/A}(s)$ is  the 1+0 spin combination for the vector/axial vector spectral function, $\Pi^{(1+0)}_{{\rm OPE},V/A}(s)$ is the OPE approximation to the corresponding correlator, and $\mbox{Im}\,
\Delta_{V/A}(s)$ contains the so-called Duality Violations (DVs) \cite{CGP}. This  DV term is supposed to compensate for the lack of convergence of the OPE on the Minkowski axis and, in more physical terms, is responsible for the mismatch between the short-distance quark-gluon description and the long-distance hadronic description of the correlator $\Pi^{(1+0)}_{{\rm OPE},V/A}(s)$. This correlator contains the perturbative series (and the nonperturbative condensates) and depends on $\alpha_s$, whereas $\rho^{(1+0)}_{V/A}(s)$ contains the experimental data. The maximum value of $s_0$ is $m_\tau^2$.

It is clear that Eq. (\ref{sumrule}) can only determine $\alpha_s$ after one assumes something about $\mbox{Im}\,
\Delta_{V/A}(s)$. The traditional method \cite{BNP,LDP}, in the most recent
incarnation of Refs. \cite{ALEPH13, Pich}, chooses the polynomial $w(s/s_0)$
from the set $w_{k\ell}(x)=(1-x)^{k+2}x^\ell(1+2x)$, with
$(k\ell)\in\{(00),(10),(11), (12),(13)\}$. The hope is that, because these
polynomials have a zero on the Minkowski axis of degree two or three, the
contribution from the OPE on the region of the contour $|s|=s_0$ touching this
axis, where the OPE is not valid,
will be suppressed enough not to require any compensating term in the form of
$\mbox{Im}\, \Delta_{V/A}(s)$. To make this possibility more likely,
$s_0$ is chosen equal to $m_\tau^2$, since the OPE works better at higher
scales. With all these choices, the net result is that
$\mbox{Im}\,\Delta_{V/A}(s)$  is assumed to be zero. However, the use of the
above set $w_{k\ell}(x)$  in Eq. (\ref{sumrule}), at the single value
$s_0=m_\tau^2$, yields only 5 data points, whereas these polynomials contain
powers of $s$ all the way up to $s^7$. Cauchy's theorem then tells us that all
condensates, $C_D$, from the OPE will contribute to Eq. (\ref{sumrule}) for
$D=4,6,8,10,12,14$ and $16$. Together with $\alpha_s$, that makes 8 unknowns
for 5 data points. In order  to be able to do a fit with one degree of
freedom, the condensates with $D=10,12,14$ and $16$ are set to zero by
hand. The bottom line is that one has to make two assumptions for the
traditional method to be operational: negligible DVs
and a truncation of higher-dimension OPE terms.
These two assumptions are actually related \cite{Peris}.

In contrast, the method of Ref. \cite{us} chooses the following  explicit parametrization for $\rho_{V/A}^{\rm DV}(s)$ $\equiv \frac{1}{\pi}\mbox{Im}\,\Delta_{V/A}(s)$:
\begin{equation}
\label{ansatz}
\rho_{V/A}^{\rm DV}(s)=
e^{-\delta_{V/A}-\gamma_{V/A}s}\sin{(\alpha_{V/A}+\beta_{V/A}s)}\ ,\qquad s\ge s_{\rm min}\ .
\end{equation}
which is based on large-$N_c$ and Regge considerations \cite{Blok}-\cite{Cata2}. For this reason we will refer to the method of Ref. \cite{us} as the "DV model" and to the method of Ref. \cite{Pich} as the "truncated-OPE (TOPE) model" .  Notice that the truncated OPE model makes the choice $\delta_{V/A} = \infty$ from the outset.

 It turns out that the DV model allows us to avoid the above OPE truncation,
 making a self-consistent fit possible in a window $s_{min} \leq s_0 \leq
 m_\tau^2$, in which  $\alpha_s(m_\tau), C_{6,8}$\footnote{$C_2$ is negligible
   and the weights are deliberately chosen to avoid $C_4$, because the
   presence of this condensate  spoils the perturbative convergence
   \cite{BBJ}.  } and the 8 DV parameters $\delta_{V/A},
 \gamma_{V/A},\alpha_{V/A}, \beta_{V/A}$ are determined from the data. The
 fits also determine $s_{min}\simeq 1.55\, \mathrm{GeV}^2$ as
an optimal choice. Consistent results were obtained in Ref. \cite{us} from a series of fits  of this type. However, the results obtained for $\alpha_s(m_\tau)$ were systematically $\sim 0.020-0.025$ lower than  in the TOPE model. It is important to clarify the underlying cause for this difference.

 Two recent analysis \cite{Pich,newpaper} have taken up this task. In
 Ref.~\cite{Pich} a very thorough survey of a variety of fits based on the
 TOPE strategy has been performed checking for the stability of the result
 with respect to the inclusion of the first higher term of the OPE
 neglected in the earlier fits. Of course, once a new term  is included,
the number of degrees of freedom is zero, the value of the fit
quality vanishes and one no longer has a real "fit". However,
errors can still be propagated, and Ref. \cite{Pich} finds results which display a certain stability with respect to this change. This stability is then taken by the authors of Ref. \cite{Pich} as a proof for the robustness and reliability of the results.

It is important to realize, however, that this condition of stability is
a necessary, but not sufficient one. It is easily demonstrated that other
choices for the higher dimension $C_D$ exist which are reasonable from the
point of view of what one could expect in QCD, but which produce a
\emph{different} result for $\alpha_s(m_\tau)$, one which is
\emph{equally stable} when tested as above \cite{newpaper}. This means
the values assumed for the higher $C_D$ affect in a significant way
the $\alpha_s(m_\tau)$ extracted from fits using the TOPE model strategy.
The tests conducted in Ref. \cite{Pich} are thus inconclusive.

 Is there a test that discriminates between the DV model strategy
and the TOPE model strategy for determining $\alpha_s(m_\tau)$?
The answer is yes. It consists of fitting a set of fake data for the $V+A$ spectral function\footnote {According to common lore, this is the combination most favorable for the TOPE model.} constructed with a known value for $\alpha_s(m_\tau)(=0.312$, using CIPT), and the DV parametrization in Eq.~(\ref{ansatz})\footnote{The precise values for the parameters can be found in Ref.~\cite{newpaper}.} by generating a multivariate gaussian distribution at the ALEPH bin energies with central values given by the above parameters and with fluctuations controlled by the real-data covariance matrix. The true and fake $V+A$ data are compared in Fig.~1. They look remarkably similar.

\begin{figure}[t]
\begin{center}
\includegraphics*[width=7.2cm]{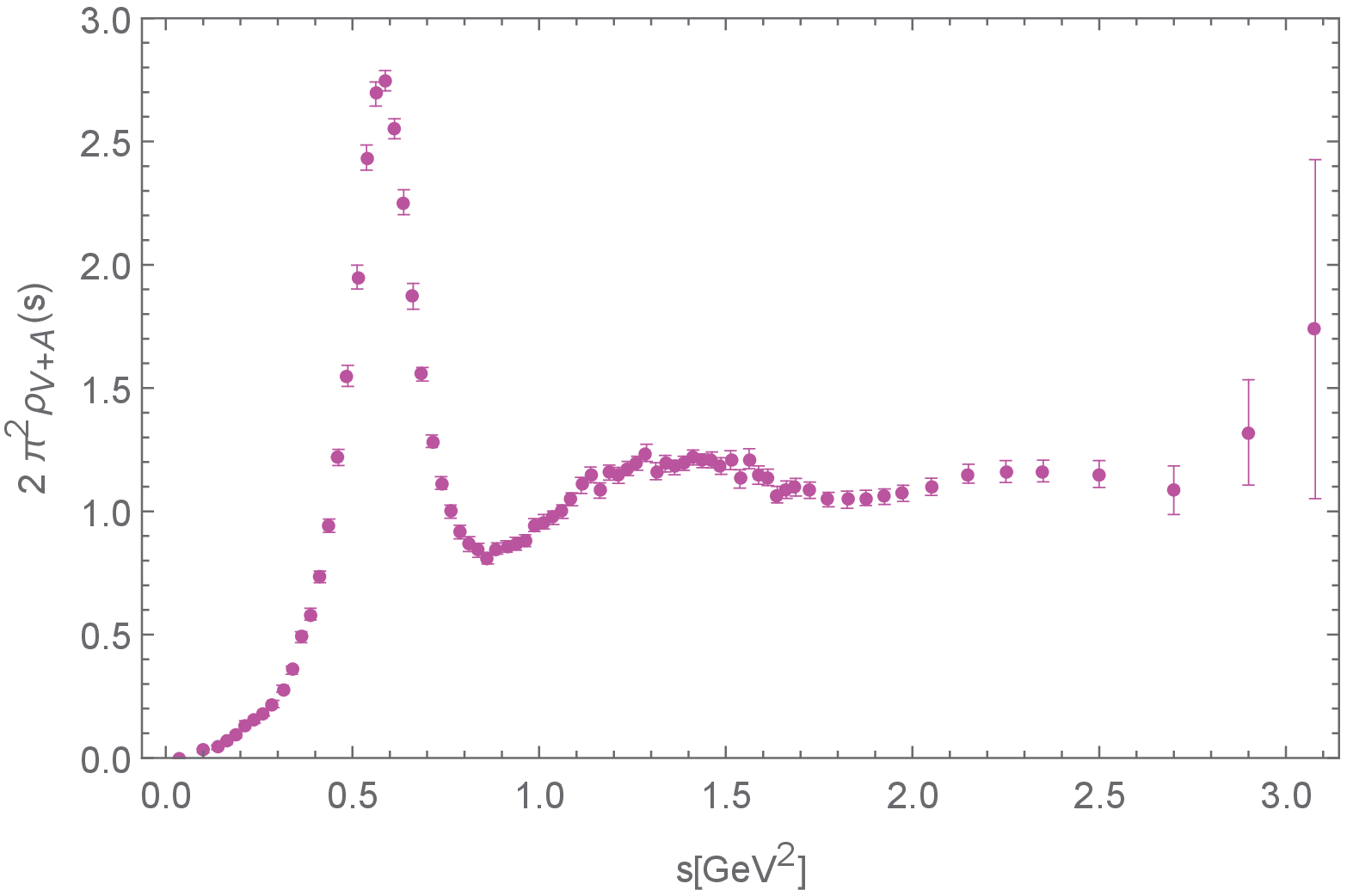}
\hspace{.1cm}
\includegraphics*[width=7.2cm]{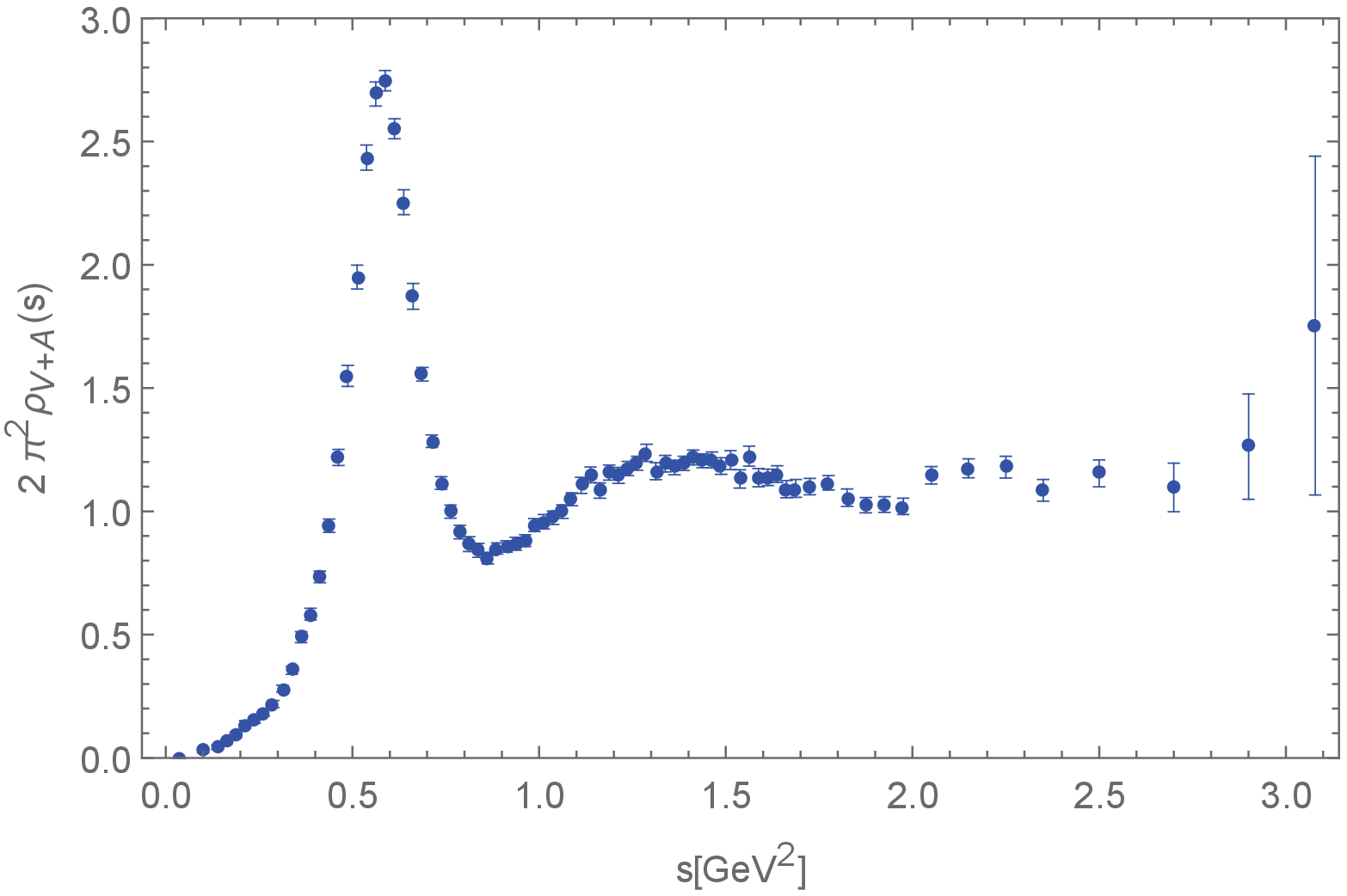}
\end{center}
\begin{quotation}
\caption{\it $V+A$ non-strange spectral function. Left panel: fake data, generated
as described in the text, as a function of $s$. Right panel:
true ALEPH data \cite{ALEPH13} as a function of $s$.
The fake data have been generated for $s\ge 1.55$~{\rm GeV}$^2$; below this
value the two data sets are the same.}
\end{quotation}
\vspace*{-4ex}
\end{figure}

If the TOPE model strategy fails to find the correct model value for $\alpha_s(m_\tau)$,
it will have been shown to be unreliable and hence not safely usable for
the real data either. This is in fact what has been found
\cite{newpaper}. The TOPE model strategy extracts a value,
$\alpha_s(m_\tau)\simeq 0.334(4)$, many standard deviations away from the
correct model result, $0.312$. The DV model strategy, on the other hand,
can easily reproduce the right value of $\alpha_s(m_\tau)$.

 One could remark that it was easy for the DV model to get the right answer
 because it was the DV model that was used as the basis for the construction
 of the fake data. While this is of course true, the above exercise shows
clearly that, by declaring DVs and the high-dimension $C_D$ to vanish,
the TOPE model leaves itself in general unable to extract the correct
value of $\alpha_s(m_\tau)$, making it unreliable.
The ALEPH data could have sizeable DVs and the TOPE model strategy would
never detect them. On the other hand, the DV model is able to detect the amount of DVs through the parametrization (\ref{ansatz}).

 \begin{figure}[t!]
\begin{center}
\includegraphics*[width=12cm]{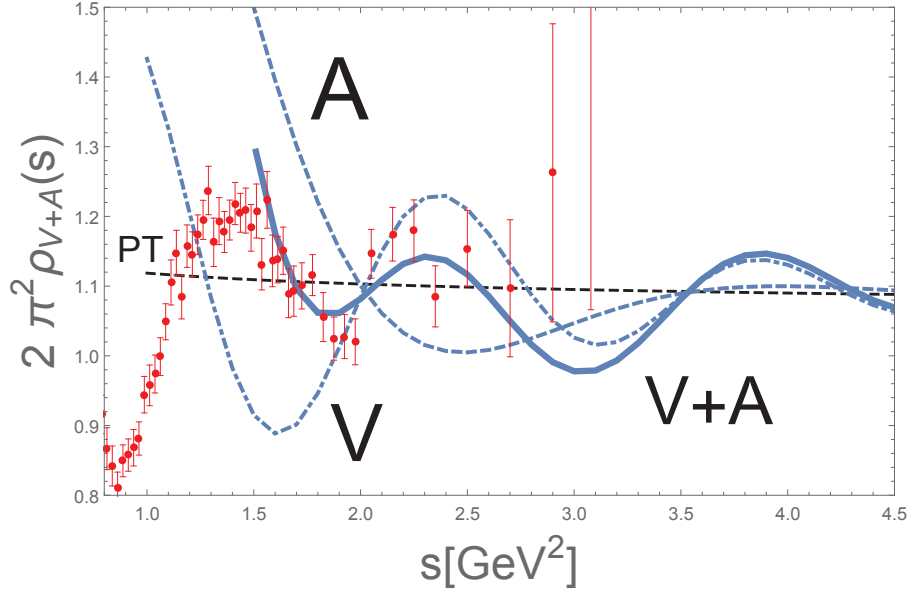}
\end{center}
\begin{quotation}
\vspace*{-2ex}
\caption{\it Blow-up of the large-$s$ region of the $V+A$ non-strange spectral function.
Black dashed line: the perturbative (CIPT) representation of
the model. Blue curve: full model representation, including DVs.
Blue dot-dashed curves: separate $V$ and $A$ parts of the model spectral
function.}
\end{quotation}
\vspace*{-4ex}
\end{figure}

 Based on our DV model fits to the ALEPH data, are the DVs large or small? Fig.~2 shows the $V+A$ ALEPH data (red points)  together with our DV model fit (solid blue line), the perturbation theory curve (in CIPT, black dashed line) and the corresponding model fits for $V$ (blue dot-dashed curve) and $A$ (blue dashed curve). As one can see, the oscillations in the data points below $s=2.3~\mathrm{GeV}^2$ are not small and show no sign of a strong damping. Above this value the data errors are too large to tell. Notice that the difference between the black dashed line (perturbation theory) and an horizontal line at $2\pi^2 \rho_{V+A}(s)=1$ (the parton model) represents the dynamical contribution from which the value of $\alpha_s(m_\tau)$ is extracted in QCD. This difference is in no way small relative to the oscillations depicted by the blue solid line (DVs) which, by the way, in the region above $s\gtrsim 1.7\ \mathrm{GeV}^2$, happen to be largest precisely at $m_\tau^2$, where the TOPE model declares them to be zero.

 In summary, the analysis of Ref.~\cite{newpaper} shows the TOPE model strategy is unreliable and, consequently, should not be used to determine $\alpha_s(m_\tau)$. The DV model strategy, on the other hand, passes all the tests known up to date\footnote{For a detailed
refutation of the criticism of our approach in Ref.~\cite{Pich}, we refer to Sec. V.B of Ref.~\cite{newpaper}.} and  obtains the following values from the ALEPH data \cite{newpaper}:

 \begin{eqnarray}
\label{finalBetal}
\alpha_s(m_\tau)&=&0.296(10)\quad (\mathrm{FOPT}) \ ,
\\
\alpha_s(m_\tau)&=&0.310(14)\quad (\mathrm{CIPT})\ .
 \end{eqnarray}

\newpage
 The work of MG is supported by the U.S. Department of Energy, Office of Science,
Office of High Energy Physics, DE-FG03-92ER40711.  The work of DB is supported by the S{\~a}o Paulo Research Foundation(FAPESP) Grant No. 2015/20689-9 and by CNPq Grant No. 305431/2015-3. KM is supported by the Natural Sciences and Engineering Research Council of Canada.  SP is supported by CICYTFEDER-FPA2014-55613-P, 2014-SGR-1450.

\end{document}